\documentclass[%
 reprint,
superscriptaddress,
 amsmath,amssymb,
 aps,
prb,
floatfix,
]{revtex4-2}

\usepackage{graphicx}
\usepackage{dcolumn}
\usepackage{bm}
\usepackage{xcolor}
\usepackage{hyperref}
\usepackage{array}
\usepackage{booktabs}

\usepackage{makecell} 

\begin{document}

\title{Non-collinear 2k antiferromagnetism in the Zintl semiconductor Eu$_5$In$_2$Sb$_6$}

\author{Vincent C. Morano}
\email{vmorano1@jhu.edu}
\affiliation{Institute for Quantum Matter and William H. Miller III Department of Physics and Astronomy, Johns Hopkins University, Baltimore, Maryland 21218, USA}
\author{Jonathan Gaudet}
\affiliation{Institute for Quantum Matter and William H. Miller III Department of Physics and Astronomy, Johns Hopkins University, Baltimore, Maryland 21218, USA}
\affiliation{NIST Center for Neutron Research, National Institute of Standards and Technology, Gaithersburg, Maryland 20899-6102, USA}
\affiliation{Department of Materials Science and Engineering, University of Maryland, College Park, MD 20742-2115, USA}
\author{Nicodemos Varnava}
\affiliation{Department of Physics and Astronomy, Rutgers University, Piscataway, New Jersey 08854, USA}
\author{Tanya Berry}
\affiliation{Institute for Quantum Matter and William H. Miller III Department of Physics and Astronomy, Johns Hopkins University, Baltimore, Maryland 21218, USA}
\affiliation{Department of Chemistry, Johns Hopkins University, Baltimore, MD 21218, USA}
\author{Thomas Halloran}
\affiliation{Institute for Quantum Matter and William H. Miller III Department of Physics and Astronomy, Johns Hopkins University, Baltimore, Maryland 21218, USA}
\author{Chris J. Lygouras}
\affiliation{Institute for Quantum Matter and William H. Miller III Department of Physics and Astronomy, Johns Hopkins University, Baltimore, Maryland 21218, USA}
\author{Xiaoping Wang}
\affiliation{Neutron Scattering Division, Oak Ridge National Laboratory, Oak Ridge, Tennessee 37831, USA}
\author{Christina M. Hoffman}
\affiliation{Neutron Scattering Division, Oak Ridge National Laboratory, Oak Ridge, Tennessee 37831, USA}
\author{Guangyong Xu}
\affiliation{NIST Center for Neutron Research, National Institute of Standards and Technology, Gaithersburg, Maryland 20899-6102, USA}
\author{Jeffrey W. Lynn}
\affiliation{NIST Center for Neutron Research, National Institute of Standards and Technology, Gaithersburg, Maryland 20899-6102, USA}
\author{Tyrel M. McQueen}
\affiliation{Institute for Quantum Matter and William H. Miller III Department of Physics and Astronomy, Johns Hopkins University, Baltimore, Maryland 21218, USA}
\affiliation{Department of Chemistry, Johns Hopkins University, Baltimore, MD 21218, USA}
\affiliation{Department of Materials Science and Engineering, Johns Hopkins University, Baltimore, Maryland 21218, USA}
\author{David Vanderbilt}
\affiliation{Department of Physics and Astronomy, Rutgers University, Piscataway, New Jersey 08854, USA}
\author{Collin L. Broholm}
\affiliation{Institute for Quantum Matter and William H. Miller III Department of Physics and Astronomy, Johns Hopkins University, Baltimore, Maryland 21218, USA}
\affiliation{NIST Center for Neutron Research, National Institute of Standards and Technology, Gaithersburg, Maryland 20899-6102, USA}
\affiliation{Department of Materials Science and Engineering, Johns Hopkins University, Baltimore, Maryland 21218, USA}

\begin{abstract}
Eu$_5$In$_2$Sb$_6$ is an orthorhombic non-symmorphic small band gap semiconductor with three distinct Eu$^{2+}$ sites and two low-temperature magnetic phase transitions. The material displays one of the greatest (negative) magnetoresistances of known stoichiometric antiferromagnets \cite{Rosa2020} and belongs to a family of Zintl materials that may host an axion insulator \cite{nico2022}. Using single crystal neutron diffraction, we show that the $T_{\mathrm{N1}}=14$~K second-order phase transition is associated with long-range antiferromagnetic order within the chemical unit cell $\left( \bm{k}_1 = (000) \right)$. Upon cooling below $T_{\mathrm{N1}}$, the relative sublattice magnetizations of this structure vary until a second-order phase transition at $T_{\mathrm{N2}}=7$~K that doubles the unit cell along the $\bm{\hat{c}}$ axis $\left( \bm{k}_2 = \left(00\frac{1}{2}\right) \right)$. We show the anisotropic susceptibility and our magnetic neutron diffraction data are consistent with magnetic structures described by the $\Gamma_3$ irreducible representation with the staggered magnetization of the $\bm{k}_1$ and $\bm{k}_2$ components polarized along the $\bm{\hat{b}}$ and $\bm{\hat{a}}$ axis, respectively. As the $\bm{k}_2$ component develops, the amplitude of the $\bm{k}_1$ component is reduced, which indicates a 2$\bm{k}$ non-collinear magnetic structure. Density functional theory is used to calculate the energies of these magnetic structures and to show the $\bm{k}_1$ phase is a metal so $T_{\mathrm{N1}}$ is a rare example of a unit-cell-preserving second-order phase transition from a paramagnetic semiconductor to an antiferromagnetic metal. DFT indicates the transition at $T_{\mathrm{N2}}$ to a doubled unit cell reduces the carrier density of the metal, which is consistent with resistivity data \cite{Rosa2020}.
\end{abstract}

\maketitle


\section{\label{sec:Intro}Introduction}
Europium Zintl materials have displayed some of the largest reported (negative) magnetoresistances of antiferromagnetic (AFM) compounds. These include 122 Eu-based antiferromagnets like EuZn$_2$As$_2$ and EuCd$_2$P$_2$ \cite{wang2022anisotropy, wang2021colossal}. Recently, the magnetoresistance of the Zintl material Eu$_5$In$_2$Sb$_6$ at 9 T and 15 K was reported to be -99.999\% \cite{Rosa2020}. Eu$_5$In$_2$Sb$_6$ also displays a record negative piezoresistance\cite{ghosh2022}. The extreme sensitivity of electronic transport to external stimuli may make Eu$_5$In$_2$Sb$_6$ useful for dark matter detection \cite{chanakian2015high, essig2022snowmass2021}. Eu$_5$In$_2$Sb$_6$ has also been proposed as an axion insulator candidate \cite{Rosa2020}. While recent work contests this hypothesis, related chemically substituted compounds may yet realize the axionic state \cite{nico2022}.

Given the rich interplay of magnetic and electronic properties, the long-range magnetic order in Eu$_5$In$_2$Sb$_6$ is of particular interest. Two magnetic phase transitions have been observed at $T_{\mathrm{N1}}=14$ K and $T_{\mathrm{N2}}=7$ K, with susceptibility data indicating antiferromagnetic structures with competing interactions \cite{Rosa2020, crivillero2023magnetic}. Here we present a single crystal neutron diffraction study of Eu$_5$In$_2$Sb$_6$. We will refer to the 0 T magnetic phase from 14 K to 7 K as ``phase 1" and the 0 T magnetic phase found from 7 K down to at least 1.5 K as ``phase 2." We identify the magnetic propagation vectors in both phases and candidate magnetic structures based on comparison of the observed and calculated structure factors for magnetic structures that can develop through a second-order phase transition. With knowledge of the ordered magnetic structure, we use DFT to calculate the electronic band structure finding that the gap in the electronic density of states within the paramagnetic (PM) phase closes in the antiferromagnetic phases.

\section{\label{sec:Meths}Methods}

Single-crystalline samples of Eu$_5$In$_2$Sb$_6$ were grown using the flux technique with InSb as the flux solvent~\cite{Rosa2020}. Powder X-ray diffraction (XRD) data, used to confirm phase purity of the single crystals, were collected over a scattering angle range of $5^{\circ}-60^{\circ}$ on a laboratory Bruker D8 Focus diffractometer with a LynxEye detector and Cu K$\alpha$ radiation.

Neutron diffraction experiments (Appendix~\ref{sec:neu}) were performed at the NIST Center for Neutron Research (NCNR) on the SPINS and BT7 \cite{lynn2012double} triple-axis spectrometers. A spherical absorption correction with effective radius corresponding to the sample volume was used to partially correct for the strong neutron absorption by Eu and In \cite{maslen2006x}. Structural and magnetic refinements were performed using FullProf \cite{zheludev2007reslib,RODRIGUEZCARVAJAL199355} with absorption corrected structure factors as the experimental input. Group theoretical analysis was performed in SARAh and ISODISTORT from the ISOTROPY software suite \cite{wills2000new,stokes,stokes2006isodisplace}. Magnetic structure factors were also calculated analytically in Mathematica \cite{Mathematica}. A single crystal neutron diffraction experiment was performed at Oak Ridge National Laboratory (ORNL) on TOPAZ. Analysis of those data was performed using the Mantid software package \cite{ARNOLD2014156, mantid5analysis}. Code and data are available on GitHub \cite{git}. Error bars represent one standard deviation unless otherwise noted.

To determine the quality and critical temperature of our sample prior to neutron diffraction, heat capacity was measured on a 5.2(1) mg single crystal cut from the BT7 sample (Fig.~\ref{fig:hc}, Fig.~\ref{fig:overplot}a). The crystal was mounted on a heat capacity puck with Apiezon N Grease for measurement in a Quantum Design PPMS using the standard heat capacity option. Heat capacity was measured using the semi-adiabatic method with a 2\% temperature rise. Fitting to the Debye model was done using MATLAB \cite{MATLAB:2023}.

Density functional theory \cite{hohenberg-pr64,kohn-pr65} (DFT) based first principles calculations were performed using the projector augmented-wave (PAW) method as implemented in the VASP code \cite{kresse-prb96,kresse-prb99}. We used the PBE exchange-correlation functional as parametrized by Perdew-Burke-Ernzerhof \cite{perdew-prl96}. The Brillouin zone sampling was performed by using a $5 \times 5 \times 15$ k-grid. The energy cutoff was chosen 1.5 times as large as the values recommended for the relevant PAW pseudopotentials. Spin-orbit coupling (SOC) was included self-consistently. The Eu $4f$ states were treated by employing the GGA+U approach with the U value set to 5.0~eV. Dipole energy calculations were performed in Mathematica \cite{Mathematica} using the phase 1 and phase 2 magnetic structures refined from neutron diffraction. The net magnetic dipole-dipole interaction strength for these antiferromagnetic structures was calculated by adding the contributions from all pairs of moments within an increasing spherical volume until convergence was achieved for a diameter of 100 \AA.

\section{\label{sec:Res}Experimental Results}

\begin{figure}
\includegraphics[scale=1.0]{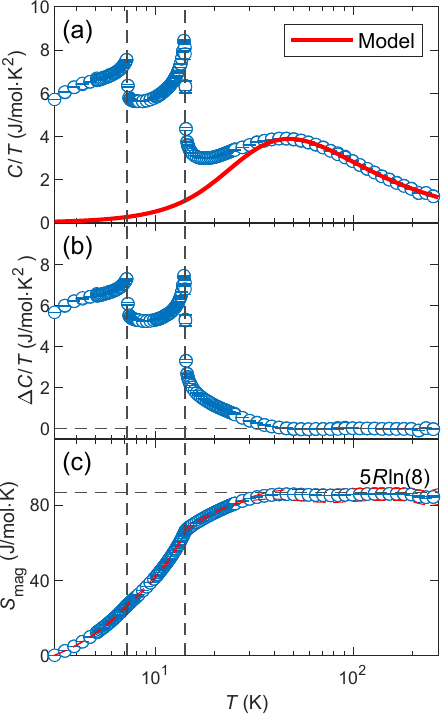}
\caption{\label{fig:hc} (a-c) Specific heat measurements on part of the $\rm Eu_5In_2Sb_6$ crystal used for neutron diffraction on BT7. (a) Debye fit with a correction for a small amount of grease being transferred during the sample mounting, indicated by the red line. $\Theta_{\rm{D}} = 170(6)\ \mathrm{K}$. We find two sharp $\lambda$ peaks near $T_{\mathrm{N1}}=14.04$~K and $T_{\mathrm{N2}}=7.20$~K (indicated with vertical dashed lines) with a broad peak at 50 K. We interpret the sharp peaks as second-order transitions. The paramagnetic phase lies above $T_{\mathrm{N1}}$, ``phase 1" is between $T_{\mathrm{N1}}$ and $T_{\mathrm{N2}}$, and ``phase 2" is below $T_{\mathrm{N2}}$. See Appendix~\ref{sec:hcapp} for an alternative fit using anharmonic contributions as in \cite{stern1958}. (b) An estimate of the magnetic heat capacity determined by subtracting our calculated Debye contribution from the observed heat capacity. The magnetic contribution only declines to 0 J/mol$\cdot$K$^2$ beyond $50$ K. (c) Magnetic entropy saturates near $5 R \log{8}$ J/mol$\cdot$K (indicated by the horizontal dashed line), the expected value for a formula unit with five $S=7/2$ Eu$^{2+}$ ions. The error bars are correlated. Dashed red lines about the calculated entropy are fits using upper and lower bounds on the measured sample mass.}
\end{figure}

Heat capacity measurements show two anomalies, neither of which have latent heat (Fig.~\ref{fig:hc}a,b). We interpret these anomalies as marking second-order phase transitions and from the midpoint of the sharp edges of the specific heat peaks we obtain $T_{\mathrm{N1}}=14.04$~K and $T_{\mathrm{N2}}=7.20$~K. Upper bounds on the half width at half maximum for the upper, sharper edges of these peak are 0.3 K and 0.1 K for $T_{\mathrm{N1}}$ and $T_{\mathrm{N2}}$, respectively. The small values of $\Delta T_+/T_{\mathrm{N1}}=0.02$ and $\Delta T_+/T_{\mathrm{N2}}=0.01$ indicate both transitions are associated with magnetic symmetry breaking in a homogeneous solid. 

\begin{figure}
\includegraphics[scale=0.5]{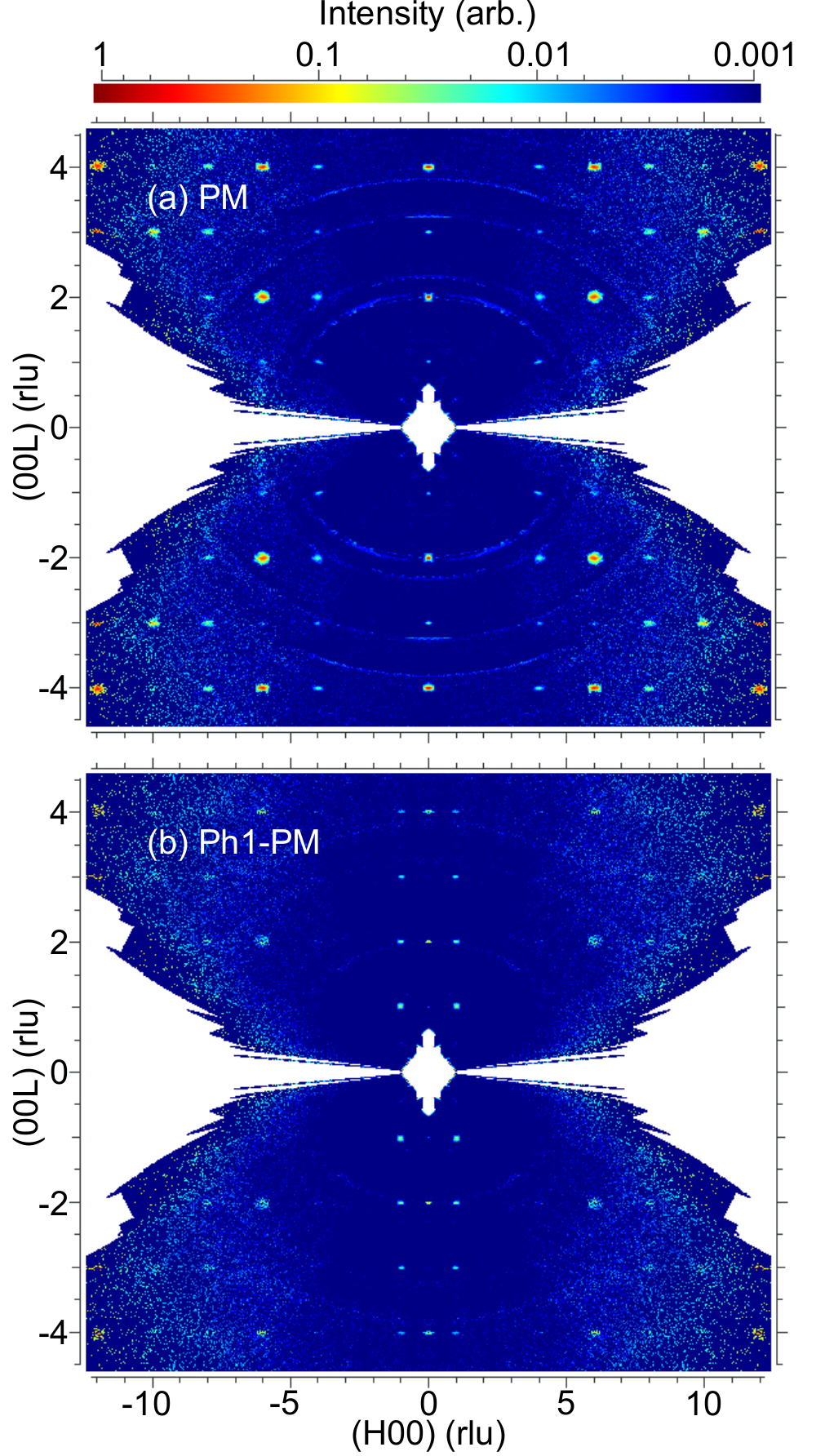}
\caption{\label{fig:topaz} (a) Intensity in the $(h0l)$ plane of the Eu$_5$In$_2$Sb$_6$ paramagnetic phase. Note that for the crystallographic space group ($Pbam$), a reflection of the form $(h0l)$ or $(0kl)$ must have $h$ or $k$ be an even integer, respectively. This reflection condition is obeyed. Some powder lines are visible. (b) Intensity difference between base temperature scans and scans in the paramagnetic phase. The four inequivalent $(10l)$ reflections are purely magnetic. We find only integer-valued peaks at base temperature in this experiment, which are associated with phase 1. Data from TOPAZ.}
\end{figure}

The corresponding magnetic Bragg peaks of phase 1 are apparent in the single crystal time of flight diffraction data from TOPAZ at ORNL \cite{coates2018suite}. Figure~\ref{fig:topaz} shows neutron scattering intensity with momentum transfer spanning the $(h0l)$ reciprocal lattice plane. In the paramagnetic phase (Fig.~\ref{fig:topaz}a) nuclear Bragg peaks consistent with the orthorhombic space group $Pbam$ (55) are observed. Figure~\ref{fig:topaz}b shows the additional Bragg diffraction that develops upon cooling into phase 1. Several observations can immediately be made: Magnetic intensity in phase 1 occurs only for integral Miller indices. This indicates the magnetic wave vector for phase 1 is $\bm{k}_1=(000)$, i.e., the magnetic unit cell equals the chemical unit cell. However, magnetic Bragg peaks appear at $(\pm 1 0 l)$ for integral $l$ where nuclear contributions are forbidden by the $(2n,0,l)$ $Pbam$ selection rule \cite{ict}. This indicates antiferromagnetic ordering of Eu spins within one or more of the three distinct Wyckoff sites. From the presence of considerable magnetic intensity on reflections such as (102), (103) and even (104) we learn that this cannot be $\bm{\hat{c}}$-polarized antiferromagnetism as magnetic Bragg diffraction is insensitive to spin components polarized along wave vector transfer. 

\begin{figure}
\includegraphics[scale=0.92]{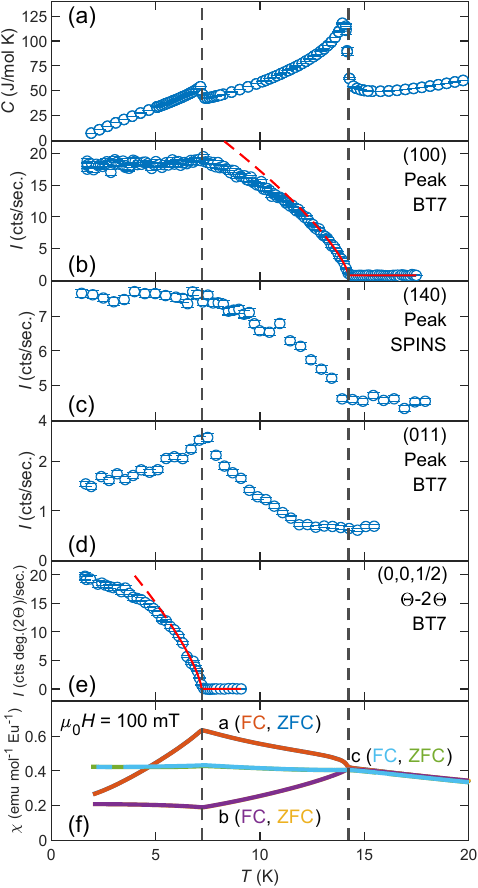}
\caption{\label{fig:overplot} (a) Eu$_5$In$_2$Sb$_6$ heat capacity with two magnetic transitions indicated by lines at $T_{\mathrm{N1}}=14.24$~K and $T_{\mathrm{N2}}=7.20$~K. The former temperature was determined by fitting to the (100) peak for $\beta$ while the latter was from heat capacity as described in Figure~\ref{fig:hc}a. (b-e) We find the same transitions in both our heat capacity and our neutron diffraction results. Forbidden integer peaks first appear in phase 1 (below $T_{\mathrm{N1}}$) while half-integer peaks emerge in phase 2 (below $T_{\mathrm{N2}}$). Note that $\bm{Q} = \left( 0 0 \frac{1}{2} \right)$ consists of rocking scans while the others have the detector stationary at the peak position. The solid red line in (b) indicates the fitted region for $\beta$. The critical behavior modeled in (e) is intended as a guide for the eye. It assumes $\beta$ for a Heisenberg class and $T_{\mathrm{N2}}$ from heat capacity. (b), (d) and (e) are from BT7 data while (c) is from SPINS. Energies were $E=14.7$ meV for (b), $E=5$ meV for (c), $E=35$ meV for (d) and $E=14.7$ meV for (e). (f) Susceptibility versus temperature measured on Eu$_5$In$_2$Sb$_6$ single crystals with a $\mu_0H=100 ~\rm{mT}$ field along $\bm{\hat{a}}$, $\bm{\hat{b}}$ and $\bm{\hat{c}}$ with masses 16.0(1) mg, 17.2(1) mg and 1.6(1) mg, respectively. Contributions to the systematic uncertainty from the mass measurements are 0.6\%, 0.6\% and 6\%, respectively. Susceptibility along $\bm{\hat{a}}$ is enhanced while susceptibility along $\bm{\hat{b}}$ is suppressed at $T_{\mathrm{N1}}$. Susceptibility along $\bm{\hat{a}}$ is suppressed at $T_{\mathrm{N2}}$. Measurements were performed on warming after cooling in zero field (ZFC) or in the measurement field (FC) as indicated in the figure.}
\end{figure}

We used triple-axis spectrometers at NIST to probe the detailed temperature dependence of magnetic diffraction. All the magnetic peaks were found to be resolution-limited indicating static long-range magnetic ordering. Temperature dependent Bragg intensity was observed at nuclear peaks such as (140) (Fig.~\ref{fig:overplot}c), which are nuclear allowed reflections for all but Eu2 on the 2a Wyckoff site.

\begin{figure}
\includegraphics[scale=1.0]{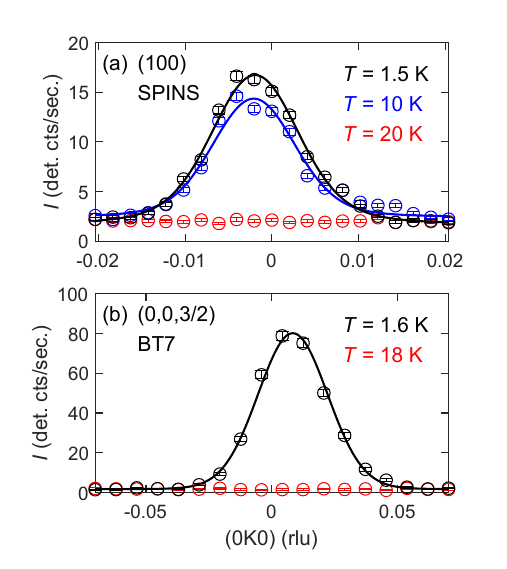}
\caption{\label{fig:peaks} (a,b) Rocking scans of Eu$_5$In$_2$Sb$_6$ magnetic reflections. A slight offset of these peaks from the listed scattering vectors is interpreted as a consequence of the alignment accuracy rather than physical phenomena such as incommensurability. Magnetic peaks were fit to a flat background and Voigt function with the reciprocal of the Lorentzian half-width half-maximum giving a spin correlation length (see Appendix~\ref{sec:corr}). The resolution width was set to the mean of fitted Gaussian widths from our brightest Bragg peaks (these were relatively constant as a function of $Q$). (a) Rocking scan through $\bm{Q} = (100)$, which violates a reflection condition for the parent space group $Pbam$. Data from SPINS experiment on an $(hk0)$-aligned sample. The fitted correlation length is greater than 960 \AA ~for an increase in $\chi_{\rm{r}}^2$ of 20\% from the minimum. (b) Data from BT7 experiment on an $(0kl)$-aligned sample. The fitted correlation length is greater than 1554 \AA ~for an increase in $\chi_{\rm{r}}^2$ of 20\%, indicating the peaks are resolution-limited.}
\end{figure}

Rocking scans acquired on SPINS through the purely magnetic $(100)$ Bragg peak are shown in Figure~\ref{fig:peaks}a. Consistent with the TOPAZ experiment and the $Pbam$ selection rules, there is no peak in the paramagnetic phase at 20 K. Bragg diffraction appears with the symmetry breaking magnetic order, growing in strength upon cooling. Considering the polarization factor, the presence of magnetic diffraction at (100) implies the staggered magnetization cannot be oriented along $\bm{\hat{a}}$. Since $\bm{\hat{c}}$-oriented moments are ruled out by the TOPAZ data, the staggered magnetization must be oriented along the $\bm{\hat{b}}$ direction. A fit of the $T-$dependent intensity in Figure~\ref{fig:overplot}b to $I(T)=I_0(1-T/T_{\mathrm{N1}})^{2 \beta}$ including data within 1.2 K of $T_{\mathrm{N1}}$\cite{Huvonen2012} yields $\beta = 0.37(2)$. This critical exponent is closest to the Heisenberg universality class ($\beta = 0.366$) but also consistent with XY criticality (0.349). 3D Ising criticality (0.326) lies just beyond our error bars. Assuming collinear antiferromagnetism (see Sec. \ref{sec:14K}), we find that order on the 2a Wyckoff sites yields Bragg diffraction on (100), (140), and (011). However, (011) does not have a conventional critical onset for $T<T_{\mathrm{N1}}$ (Fig.~\ref{fig:overplot}d). This implies collinear AFM ordering on the 2a position cannot be the principal order parameter of the $\bm{k}_1=(000)$ transition. 

The reduction in (011) intensity for $T<T_{\mathrm{N2}}$ should be matched with an increase in intensity elsewhere. Triple-axis scans using BT7 through the $(h0l)$ plane Brillouin zone at 1.5 K revealed half-integer peaks, which indicate the development of a new magnetic order parameter with a characteristic wave vector $\bm{k}_2=\left(00\frac{1}{2}\right)$ (Fig.~\ref{fig:overplot}e, Fig.~\ref{fig:peaks}b). The corresponding magnetic Bragg peaks are again resolution-limited. The temperature dependence of the $\left( 0 0 \frac{1}{2} \right)$ reflection near $T_{\mathrm{N2}}$ is consistent with a 3D Heisenberg criticality (dashed line in Fig.~\ref{fig:overplot}e) though the data quality does not allow for a unique determination of the universality class.

\section{\label{sec:Analysis}Analysis and Discussion}

\subsection{\label{sec:Interp}Interpretation of Magnetic Results}

The low symmetry of Eu$_5$In$_2$Sb$_6$, the strong neutron absorption of Eu, the presence of three Eu Wyckoff sites and of two magnetic phase transitions make it difficult to determine the magnetic structure of this compound by neutron diffraction alone. To proceed, we shall combine information from susceptibility measurements (\cite{Rosa2020}, Fig.~\ref{fig:overplot}f) with \textit{T}-dependent Bragg diffraction (Fig.~\ref{fig:overplot}b-e) and conventional refinement of neutron diffraction data (Fig.~\ref{fig:refinement}). As we shall see, this leads to hypothesized magnetic structures for phases 1 and 2 that are consistent with the data.

Fits of the measured heat capacity to the Debye interpolation scheme give an entropy approaching the expected $5 R \log 8$ J/mol$\cdot$K for a system with five $S=\frac{7}{2}$ Eu$^{2+}$ atoms per formula unit (Fig.~\ref{fig:hc}c). Furthermore, the anomalies at each of the phase transitions are $\lambda-$ like and carry no latent heat. We thus consider each Eu site to form a 7 $\mu_{\mathrm{B}}$ local moment and take both transitions to be second-order such that the 7 K transition simply adds the $\bm{k}_2=\left(00\frac{1}{2}\right)$ propagation vector to phase 1 resulting in a homogeneous magnetically ordered crystal with two magnetic wave vectors (see Appendix~\ref{sec:het}).

\subsection{\label{sec:14K}The 14 K Phase Transition}

The magnetic susceptibility of Eu$_5$In$_2$Sb$_6$ is highly anisotropic. The reduction in $\chi_{\rm{b}}$ upon cooling below $T_{\mathrm{N1}}$ suggests an antiferromagnetic component of the magnetic moment along $\bm{\hat{b}}$. $\chi_{\rm{c}}$ on the other hand is nearly temperature-independent across the transition indicating no component of the magnetic order lies along $\bm{\hat{c}}$. This is consistent with the TOPAZ and BT7 data that show the diffraction in the high-$T$ phase is associated with the $\bm{\hat{b}}$-component of Eu1 and/or Eu3, which occupy the 4g Wyckoff sites (Fig.~\ref{fig:topaz}b, Fig.~\ref{fig:peaks}a). $\chi_{\rm{a}}$ is enhanced and hysteresis in $M$ versus $H$ scans at 10 K indicates a very small ferromagnetic component along $\bm{\hat{a}}$ (\cite{Rosa2020}, 0.021(1) $\mu_\mathrm{B}/\mathrm{Eu}$ in Fig.~\ref{fig:MvH}).

Neutron diffraction shows the $T_{\mathrm{N1}}=14$ K transition leads to AFM order within the chemical unit cell, i.e., $\bm{k}_1=(000)$. With this propagation vector (as well as the reported parent space group and Eu positions \cite{Park2002}), we shall use the Landau theory of second-order phase transitions to narrow down the list of possible magnetic structures \cite{izyumov1980neutron}. The basis vectors of the irreducible representations of the $\bm{k}_1=(000)$ little group $G_{{\bm{k}}1}$ are given in Table~\ref{tab:irreps}.

\begin{table*}
\caption{\label{tab:irreps} The $\bm{k}_1=(000)$ and $\bm{k}_2=\left(00\frac{1}{2}\right)$ normalized basis vectors for space group $Pbam$ (55). The basis vectors are lists of (unit) vectors for each Eu atom that specify possible magnetic moment orientations. Eu1 and Eu3 both occupy 4g Wyckoff sites while Eu2 is on the 2a site. Each irreducible representation is denoted as $\Gamma_i$ where $i$ labels the irrep. Each Wyckoff site has its own set of basis vectors $\psi_j(\Gamma_i)$ and a corresponding set of mixing coefficients in $\mu_{\mathrm{B}}$. Only spin structures realizing irreps 2, 3, 5 and 8 lie in the ab-plane. Of these, only irreps 3 and 5 have all sites magnetized.}
\begin{ruledtabular}
\begin{tabular}{|*{14}{r|}}
Eu & Location & $\Gamma_1$ & \multicolumn{2}{c|}{$\Gamma_2$} & \multicolumn{2}{c|}{$\Gamma_3$} & $\Gamma_4$ & \multicolumn{2}{c|}{$\Gamma_5$} & $\Gamma_6$ & $\Gamma_7$ & \multicolumn{2}{c|}{$\Gamma_8$}\\
\midrule
- & - & $\bm{\psi}_1(\Gamma_1)$ & $\bm{\psi}_1(\Gamma_2)$ & $\bm{\psi}_2(\Gamma_2)$ & $\bm{\psi}_1(\Gamma_3)$ & $\bm{\psi}_2(\Gamma_3)$ & $\bm{\psi}_1(\Gamma_4)$ & $\bm{\psi}_1(\Gamma_5)$ & $\bm{\psi}_2(\Gamma_5)$ & $\bm{\psi}_1(\Gamma_6)$ & $\bm{\psi}_1(\Gamma_7)$ & $\bm{\psi}_1(\Gamma_8)$ & $\bm{\psi}_2(\Gamma_8)$ \\
\midrule
1,3 & $(x,y,0)$ & $\bm{\hat{c}}$ & $\bm{\hat{a}}$ & $\bm{\hat{b}}$ & $\bm{\hat{a}}$ & $\bm{\hat{b}}$ & $\bm{\hat{c}}$ & $\bm{\hat{a}}$ & $\bm{\hat{b}}$ & $\bm{\hat{c}}$ & $\bm{\hat{c}}$ & $\bm{\hat{a}}$ & $\bm{\hat{b}}$ \\
1,3 & $\left( x+\frac{1}{2},\bar{y}+\frac{1}{2},0 \right)$ & $-\bm{\hat{c}}$ & $\bm{\hat{a}}$ & $-\bm{\hat{b}}$ & $\bm{\hat{a}}$ & $-\bm{\hat{b}}$ & $-\bm{\hat{c}}$ & $-\bm{\hat{a}}$ & $\bm{\hat{b}}$ & $\bm{\hat{c}}$ & $\bm{\hat{c}}$ & $-\bm{\hat{a}}$ & $\bm{\hat{b}}$ \\
1,3 & $\left( \bar{x}+\frac{1}{2},y+\frac{1}{2},0 \right)$ & $-\bm{\hat{c}}$ & $-\bm{\hat{a}}$ & $\bm{\hat{b}}$ & $\bm{\hat{a}}$ & $-\bm{\hat{b}}$ & $\bm{\hat{c}}$ & $-\bm{\hat{a}}$ & $\bm{\hat{b}}$ & $-\bm{\hat{c}}$ & $\bm{\hat{c}}$ & $\bm{\hat{a}}$ & $-\bm{\hat{b}}$ \\
1,3 & $(\bar{x},\bar{y},0)$ & $\bm{\hat{c}}$ & $-\bm{\hat{a}}$ & $-\bm{\hat{b}}$ & $\bm{\hat{a}}$ & $\bm{\hat{b}}$ & $-\bm{\hat{c}}$ & $\bm{\hat{a}}$ & $\bm{\hat{b}}$ & $-\bm{\hat{c}}$ & $\bm{\hat{c}}$ & $-\bm{\hat{a}}$ & $-\bm{\hat{b}}$ \\
2 & $(0,0,0)$ & $\bm{\hat{c}}$ & - & - & $\bm{\hat{a}}$ & $\bm{\hat{b}}$ & - & $\bm{\hat{a}}$ & $\bm{\hat{b}}$ & - & $\bm{\hat{c}}$ & - & - \\
2 & $\left( \frac{1}{2},\frac{1}{2},0 \right)$ & $-\bm{\hat{c}}$ & - & - & $\bm{\hat{a}}$ & $-\bm{\hat{b}}$ & - & $-\bm{\hat{a}}$ & $\bm{\hat{b}}$ & - & $\bm{\hat{c}}$ & - & - \\
\end{tabular}
\end{ruledtabular}
\end{table*}
Irreps 2, 3, 5 and 8 describe magnetic order with spins in the ab-plane. Only irrep 3 allows for ferromagnetism along $\bm{\hat{a}}$. If the magnetization data for field along $\bm{\hat{a}}$ reflect an intrinsic ferromagnetic component of the order parameter, then the 14 K magnetic transition must proceed through irrep 3 and the staggered magnetization must be oriented along $\bm{\hat{b}}$. Note that if the FM component along $\bm{\hat{a}}$ is not intrinsic, perhaps indicating a 0.3\% molar fraction EuO impurity phase, then irreps 2, 5 and 8 are also viable.

Further constraints on the magnetic order are obtained by considering the $T-$dependent Bragg diffraction in Figure~\ref{fig:overplot}. There are only three spin configurations that produce magnetic Bragg diffraction at (100) and (140) but {\em not} at (011), consistent with the different temperature-dependences of the magnetic Bragg diffraction intensities at these reciprocal lattice points (Fig.~\ref{fig:overplot}b-d). These spin configurations are Eu1 (4g) ordering through $\Gamma_2$ and Eu3 (4g) ordering through $\Gamma_2$ or $\Gamma_3$. But because magnetic diffraction at (011) does emerge within phase 1 upon cooling (Fig.~\ref{fig:overplot}d), the irrep characterizing phase 1 must allow other Eu sites to produce magnetic Bragg diffraction at (011) (see Appendix~\ref{sec:land}). This is not the case for $\Gamma_2$, which thus is not a viable irrep for phase 1. We may conclude that magnetic order on the Eu3 site described by $\Gamma_3$ becomes critical at $T_{\mathrm{N1}}$ and subsequently induces $\Gamma_3$ order on the Eu1 and/or Eu2 sites. This implies $\bm{\hat{a}}$-oriented uniform magnetization for $T<T_{\mathrm{N1}}$ as is indeed observed (Fig.~\ref{fig:overplot}f, Fig.~\ref{fig:MvH}).

\begin{figure}
\includegraphics[scale=0.9]{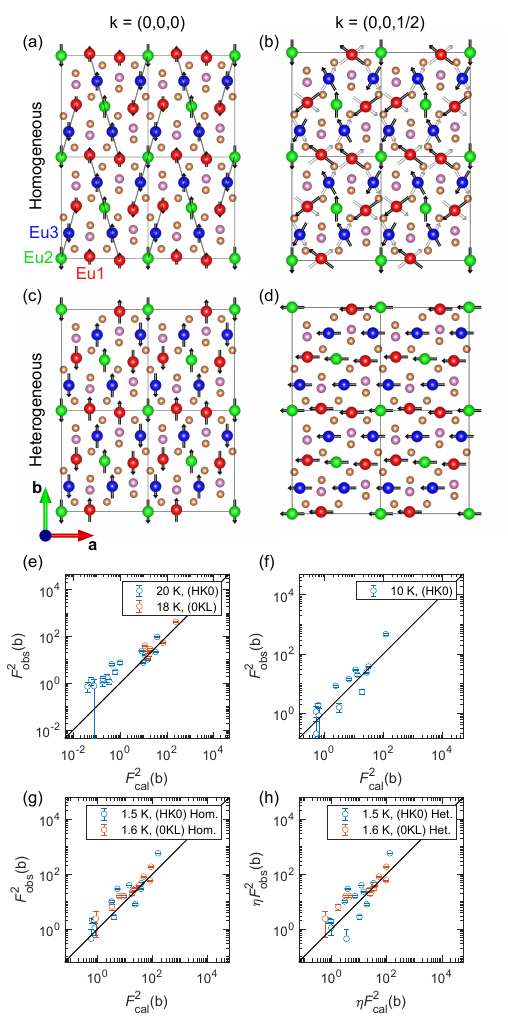}
\caption{\label{fig:refinement} (a-h) Magnetic refinements following neutron diffraction on Eu$_5$In$_2$Sb$_6$ single crystals. (a) 10 K (phase 1) homogeneous $\bm{k}_1=\left(000\right)$ refinement. Eu1 (4g) are red, Eu2 (2a) are green, and Eu3 (4g) are blue. Approximately ferromagnetic 5-atom ``rods" of Eu are centered about the Eu2 sites. (b) 1.5 K (phase 2) homogeneous $\bm{k}_1=\left(000\right)$ and $\bm{k}_2=\left(00\frac{1}{2}\right)$. Moments in black correspond to ions at $z=c$ while moments in gray are at $z=0$. (c) 1.5 K heterogeneous $\bm{k}_1=\left(000\right)$. (d) 1.5 K heterogeneous $\bm{k}_2=\left(00\frac{1}{2}\right)$. A single layer of the unit cell at constant $z$ is shown, the moments flip upon moving up or down one lattice parameter $c$. (e-f) Observed versus calculated structure factors, defined to allow negative values. NB negative values are not visible on the log-scale. (e) Paramagnetic refinement. (f) 10 K homogeneous $\bm{k}_1=\left(000\right)$. Nuclear contributions to $\bm{k}_1=\left(000\right)$ magnetic peaks were subtracted off using the PM dataset. (g) 1.5 K homogeneous $\bm{k}_1=\left(000\right)$ and $\bm{k}_2=\left(00\frac{1}{2}\right)$. Nuclear contributions to $\bm{k}_1=\left(000\right)$ magnetic peaks were subtracted off using the PM dataset. (h) 1.5 K heterogeneous phases 1 and 2 ($\bm{k}_1=\left(000\right)$ and $\bm{k}_2=\left(00\frac{1}{2}\right)$). $\eta$ indicates the refined volume fraction.}
\end{figure}

\begin{table}
\caption{\label{tab:refSum} Refined moment sizes on Eu sites 1 (4g), 2 (2a) and 3 (4g) assuming a homogeneous magnetic phase below 2 K. Errors were determined by varying one moment while refining all others until a threshold $\chi_{\rm{r}}^2$ 20\% larger than the best $\chi_{\rm{r}}^2$ was exceeded. The error bars on the refined moment size are unusually large as a result of the strong neutron absorption by Eu. Refer to $\Gamma_3$ in Table~\ref{tab:irreps} for the basis vectors of these coefficients.}
\begin{ruledtabular}
\begin{tabular}{|c | c | c || c | c | c|}
\textrm{\textit{T} (K)}&
\textrm{$\bm{k}$}&
\textrm{$\bm{\hat{\mu}}$}&
\textrm{$\mu_1 (\mu_\mathrm{B})$}&
\textrm{$\mu_2 (\mu_\mathrm{B})$}&
\textrm{$\mu_3 (\mu_\mathrm{B})$}\\
\colrule
10 & (000) & $\bm{\hat{b}}$ & 5(2) & -7(3) & -6(2) \\ 
\hline
1.6 & $\left(0 0 \frac{1}{2} \right)$ & $\bm{\hat{a}}$ & $>$-3 & $\leq$7 & 4(3) \\
\hline
1.5 & (000) & $\bm{\hat{b}}$ & 5(3) & -7(4) & -7(2) \\
\end{tabular}
\end{ruledtabular}
\end{table}
 
Also in support of this inference, refinement of the magnetic Bragg intensities in the $(hk0)$ scattering plane at 10 K using irreps 2, 3, 5 and 8 favors $\Gamma_3$ (Fig.~\ref{fig:refinement}a, see Appendix~\ref{sec:neu}). Only the AFM $\bm{\hat{b}}$ component was refined because the FM $\bm{\hat{a}}$ component is constrained by magnetization data. At 10 K, the refined structure is composed of FM 5-atom arrays (indicated by the black lines) consisting of two each Eu1 and Eu3 sites and one Eu2 site (Fig.~\ref{fig:refinement}a,e,f). Neighboring 5-atom arrays are in turn AFM correlated. Within the large error bars resulting from absorption corrections, all Eu sites carry similar ordered moments at 10 K though criticality at $T_{\mathrm{N1}}$ is driven by $\Gamma_3$ AFM order on the Eu3 site. 

\subsection{\label{sec:7K}The 7 K Phase Transition}

Turning now to the lower $T$ transition, we note that $\chi_{\rm{a}}$ is suppressed below 7 K. This indicates a component of the low-temperature antiferromagnetic order lies along $\bm{\hat{a}}$. $\chi_{\rm{b}}$ and $\chi_{\rm{c}}$, on the other hand, do not change significantly at this transition. Given the second-order nature of both phase transitions, we consider homogeneous 2k structures. Since $\bm{k}_2=\left(00\frac{1}{2}\right)$, the magnetic unit cell is doubled along $\bm{\hat{c}}$ at this transition. So while a small FM contribution along $\bm{\hat{a}}$ is possible for the $\bm{k}_1=(000)$ order, the additional magnetic order that develops below 7 K is entirely AFM. Barring strong quantum fluctuations in this 3D local moment magnet, all spins in Eu$_5$In$_2$Sb$_6$ must eventually acquire the full 7 $\mu_\mathrm{B}$ ordered moment associated with the half filled $4f$ shell. This implies the $\bm{k}_1=(000)$ and the $\bm{k}_2=\left(00\frac{1}{2}\right)$ components of the magnetic order on every Wyckoff site must have orthogonal polarizations. Were this {\em not} the case, the moment size would alternate with translations along $\bm{\hat{c}}$. As we have shown the AFM order in the $\bm{k}_1=(000)$ structure involves all three Wyckoff sites and is polarized along $\bm{\hat{b}}$, the $\bm{k}_2=\left(00\frac{1}{2}\right)$ staggered moment must be primarily oriented along $\bm{\hat{a}}$. Since Eu3 drives $T_{\mathrm{N1}}$, we expect Eu1 and/or Eu2 drive the 7 K transition, which is consistent with the suppression of the (011) intensity at $T_{\mathrm{N2}}$ (Fig.~\ref{fig:overplot}d). 

Even though the 7 K transition proceeds from a black-white group rather than a paramagnetic (grey) group, it can be shown that an arbitrary black-white group is isomorphic to a paramagnetic group and their unitary irreducible representations coincide \cite[p. 36-37]{naish2012neutron}. Specifically, our proposed $Pb^\prime am^\prime$ magnetic space group is isomorphic to the auxiliary crystallographic space group $Pbam$, which is also the parent space group. Therefore we simply consider irreps for space group $Pbam$ and $\bm{k}_2=\left(00\frac{1}{2}\right)$. For this orthorhombic structure the irreps are the same as for $\bm{k}_1=(000)$, and are listed in Table~\ref{tab:irreps}. Of these, only irreps 2, 3, 5 and 8 involve a staggered moment along the $\bm{\hat{a}}$ axis. Refinement at 1.5 K of integer-valued $(hk0)$ peaks enhanced by the $\bm{k}_1=(000)$ contribution points to irrep 3 as the best-fit (Fig.~\ref{fig:refinement}g). This is in agreement with the 10 K refinement, as the $\bm{k}_1=(000)$ irrep should not change at the $T_{\mathrm{N2}}=7$ K transition which is driven by the $\bm{k}_2=\left(00\frac{1}{2}\right)$ magnetic order. Based on our earlier analysis, we constrain the $\bm{k}_2=\left(00\frac{1}{2}\right)$ moment to (1) lie along $\bm{\hat{a}}$ and (2) to give a net moment of $7\ \mu_\mathrm{B}$ on every Eu$^{2+}$ ion up to the error bar implied by the $\bm{k}_1=(000)$ refinement. This second constraint is lifted when calculating the uncertainty in the refined moments. The structure factor is approximately dependent on the difference between the Eu1 and Eu2 moments, giving large uncertainties. A comparison of the simulated and observed $T=1.5$ K, $\bm{k}_2=\left(00\frac{1}{2}\right)$ structure factors points to $\Gamma_3$ as the best-fit solution for the 7 K transition (Fig.~\ref{fig:refinement}g). We note that a 7 K transition through irrep 3 is consistent with our temperature scan of the $\bm{Q}=\left(0 0 \frac{1}{2}\right)$ peak. Only irreps 3, 5 and 7 can account for that reflection. For irrep 7, only the $\bm{\hat{c}}$ component of the moment contributes to the reflection. But this is excluded by the polarization factor as it lies parallel to the scattering vector. Irrep 7 is furthermore inconsistent with our earlier conclusion from susceptibility measurements that staggered magnetization lies in the ab-plane. For irrep 5, only the $\bm{\hat{b}}$ component of the moment contributes to $\left(0 0 \frac{1}{2}\right)$ but this is contrary to our earlier conclusion that the staggered magnetizations associated with $\bm{k}_1$ and $\bm{k}_2$ must be perpendicular to each other. 

Thus $\Gamma_3$ is the only viable option for phase 2. Both the high- and low-$T$ magnetic orders are given by the magnetic space group 55.358 in the Belov-Neronova-Smirnova (BNS) notation, in which the unit cell is generally not the paramagnetic unit cell \cite{stokes}. This is 55.6.446 according to the Opechowski-Guccione (OG) notation, in which the unit cell is the paramagnetic unit cell. Refined moments are reported in Table~\ref{tab:refSum}. The refined 10 K homogeneous structure is shown in Figure~\ref{fig:refinement}a and the 1.5 K structure is in Figure~\ref{fig:refinement}b.

\subsection{\label{sec:dft}DFT calculations}

\begin{figure}
\includegraphics[scale=0.49]{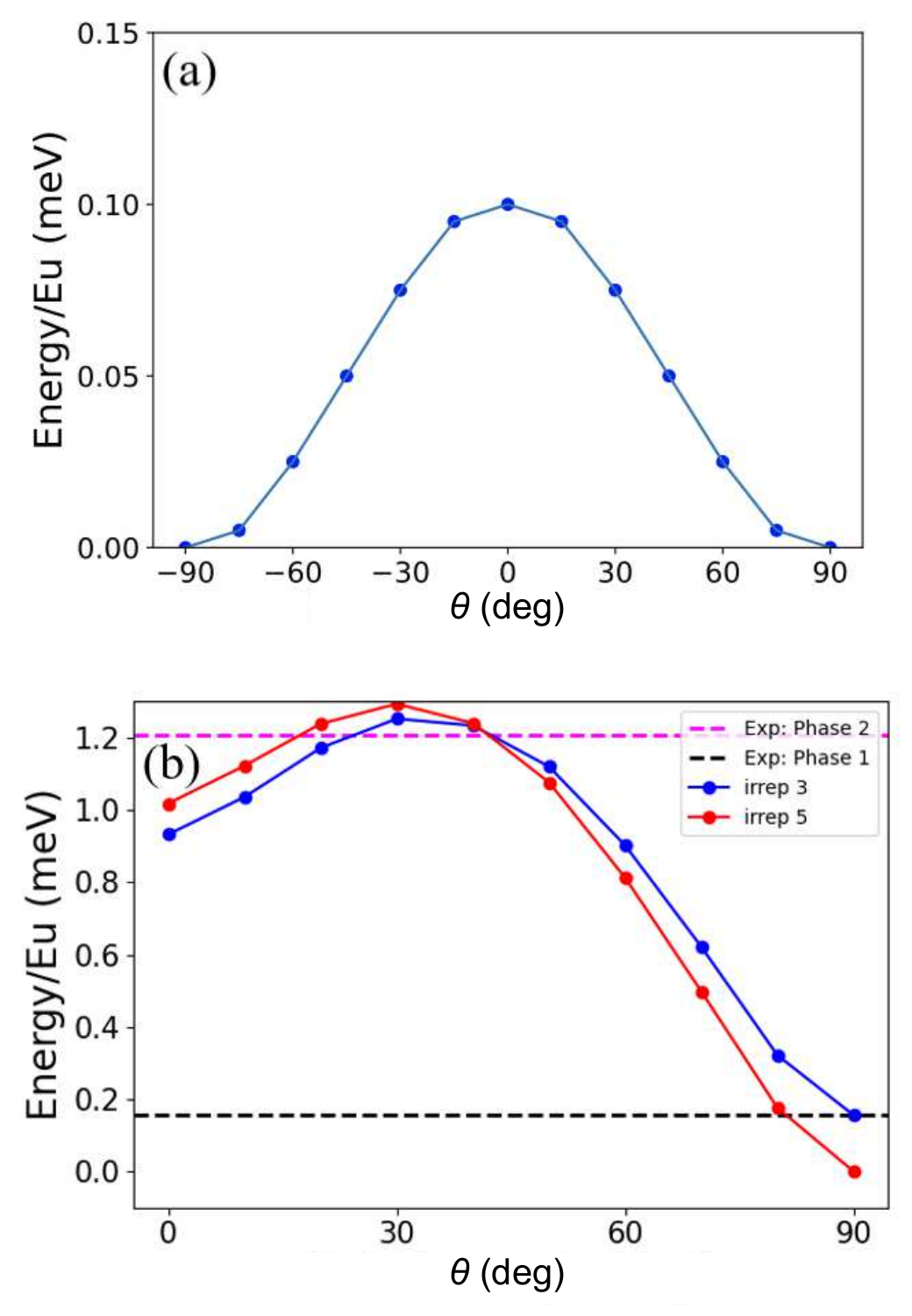}
\caption{\label{fig:energyscales} (a,b) Comparison of the total calculated energy between different Eu$_5$In$_2$Sb$_6$ magnetic configurations. (a) Estimation the magnetic anisotropy scale by uniformly rotating the spins of a reference magnetic configuration. (b) Interpolation between the low energy rod-like configurations. The black dashed line corresponds to the magnetic phase 1 shown in Fig.~\ref{fig:refinement}a while the magenta dashed line corresponds to magnetic phase 2 shown in Fig.~\ref{fig:refinement}b.}
\end{figure}

We now turn to first principles calculations that explore the DFT ground state energy of magnetic structures in $\rm Eu_5In_2Sb_6$ for comparison to the experimentally determined low$-T$ $\Gamma_3$ type order. For that structure we also determine the electronic band structure for insights into the electronic transport anomalies of the material. 

Because the susceptibility data indicate that no component of the magnetic order lies along $\bm{\hat{c}}$, we consider only in-plane magnetic configurations. First we estimate the energy scale for the in-plane magnetic anisotropy. We consider a reference magnetic configuration, in this case a ferromagnetic configuration with spins pointing along $\bm{\hat{b}}$, and then rotate all spins uniformly while monitoring the total energy. The result in Figure~\ref{fig:energyscales}a shows that FM order ha the lowest energy when oriented along $\bm{\hat{a}}$ and the scale of this anisotropy is just 0.1 meV; an order of magnitude less than $k_{\mathrm{B}}T_{\mathrm{N1}}$. 

\begin{figure*}
\includegraphics[scale=0.55]{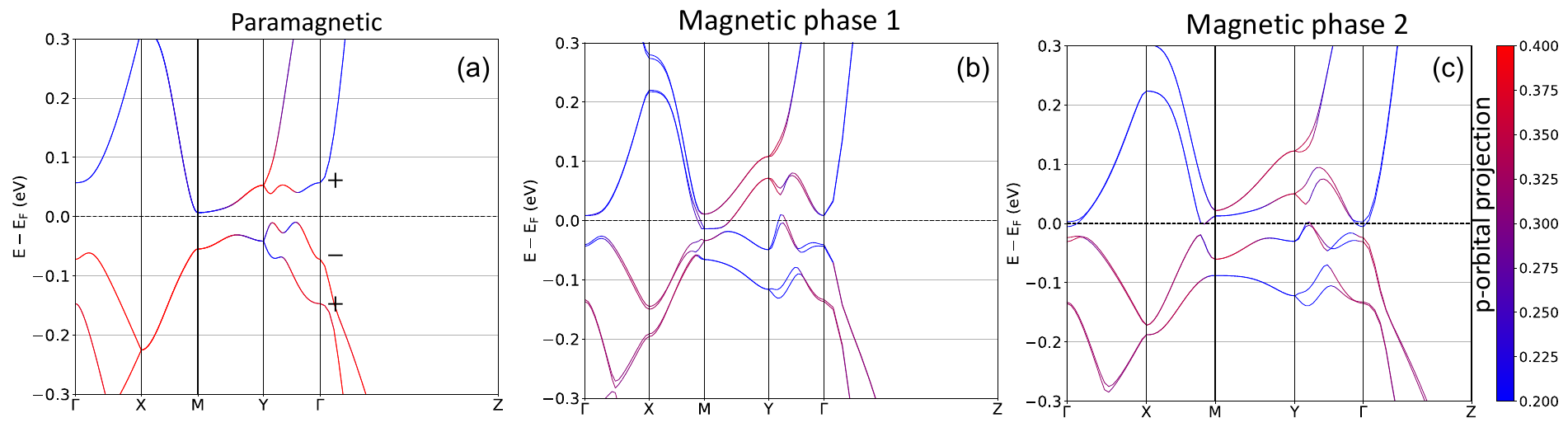}
\caption{\label{fig:bandgap} (a-c) Calculation of the Eu$_5$In$_2$Sb$_6$ band structure in each magnetic phase consistent with the diffraction data. (a) Paramagnetic state. Signs indicate the parity of each band at the $\Gamma$ point. Contrary to \cite{Rosa2020}, we find the bands are topologically trivial \cite{nico2022}. (b) Phase 1 (Fig.~\ref{fig:refinement}a). The gap perhaps closes upon magnetic ordering though the bands remain separated. (c) Phase 2 (Fig.~\ref{fig:refinement}b). We note that \cite{Rosa2020} reports an increase in resistivity with cooling into phase 2.}
\end{figure*}

Then we continuously interpolate between the lowest energy configurations. Note that all follow the rod-like geometry shown in Figure~\ref{fig:refinement} where the spins are parallel along the rod defined by the five Eu atoms. Specifically, starting from the A-type AFM configuration, which contains four rods in the magnetic unit cell, we rotate two of the rods clockwise and the other two anticlockwise so that at $90^\circ$ we get phase 1. For $45^\circ$ we approximately get phase 2 (Fig.~\ref{fig:refinement}b). This corresponds to taking the $\Gamma_3$ irrep for every Wykcoff position and both $\bm{k}$-vectors, with $7~\mu_{\mathrm{B}}$ for all Eu sites. For Eu1, Eu2, and Eu3, the $\bm{k}_2=\left(00\frac{1}{2}\right)$ components are given by $\Gamma_3(7\cos{\theta}~\mu_{\mathrm{B}},0)$, $\Gamma_3(7\cos{\theta}~\mu_{\mathrm{B}},0)$, and $\Gamma_3(7\cos{\theta}~\mu_{\mathrm{B}},0)$ while the $\bm{k}_1=(000)$ components are given by $\Gamma_3(0,7\sin{\theta}~\mu_{\mathrm{B}})$, $\Gamma_3(0,-7\sin{\theta}~\mu_{\mathrm{B}})$, and $\Gamma_3(0,-7\sin{\theta}~\mu_{\mathrm{B}})$, where $\theta$ is the rotation angle and $(\eta_1,\eta_2)$ define mixing coefficients along $\bm{\hat{a}}$ and $\bm{\hat{b}}$, respectively. However, our first principles calculations find the energy of the structure that describes our lowest-temperature diffraction data to be of a higher energy than that of the structure describing the higher-temperature diffraction data.

While dipole interactions were not included in the DFT, we calculated the energy per Eu atom associated with dipole interactions up to some distance $d_{\textrm{max}}$ from each site in the unit cell. After reaching convergence, assuming homogeneous magnetic order with refined moments from our neutron diffraction, the dipole energy of phase 1 is -0.07 meV and the dipole energy of phase 2 is -0.14 meV. These energies (and those of other tested configurations) are $~10\%$ smaller than the energy separation between phase 1 and phase 2 determined by DFT, therefore the discrepancy is not due to dipole interactions. Further work will be required to understand this discrepancy between theory and experiment. Uncertainty in the chemical potential, however, may dramatically change the Fermi surface. Given the experimental challenges stated above, it is likely that the refined moment orientations deviate somewhat from the ideal structure. The refined moments from diffraction have uncertainties that, despite not significantly affecting the quality of the refinement, may yield significantly different energies in DFT.

We report band structures in each of the three magnetic phases that are consistent with the diffraction data (Fig.~\ref{fig:bandgap}). These calculations were performed for a refinement uncorrected for absorption (see Appendix~\ref{sec:neu}). While the band gaps should be treated with caution, we qualitatively find the paramagnetic phase is a narrow gap semiconductor (Fig.~\ref{fig:bandgap}a). With long-range magnetic order at 14 K, the band gap closes and Eu$_5$In$_2$Sb$_6$ becomes metallic (Fig.~\ref{fig:bandgap}b). The gap remains closed upon cooling but perhaps begins to reopen below 7 K. The influence of magnetic order on the band structure of this narrow gap semiconductor may explain interesting changes in resistivity, including the large decrease upon antiferromagnetic ordering at 14 K \cite{Rosa2020}.

\section{\label{sec:Concl}Conclusion}

The combination of magnetization data, temperature-dependent magnetic diffraction data, and Rietveld analysis of magnetic diffraction data at 10 K and 1.5 K leads us to the following tentative conclusion about the magnetic structure of Eu$_5$In$_2$Sb$_6$.

The 14 K transition is into a $\bm{k}_1=(000)$ antiferromagnet described by irrep 3 with a very small ferromagnetic moment along the $\bm{\hat{a}}$ axis and an antiferromagnetic moment along $\bm{\hat{b}}$. This $\bm{k}_1=(000)$ component of the magnetic structure is identical in all ab-planes. The principal order parameter is at the Eu3 site while magnetization is induced on the Eu1 and/or Eu2 sites upon cooling. We find a phase 1 magnetic structure in which approximately linear arrangements of five near-neighbor Eu atoms (consisting of all three Wyckoff sites) adopt ferromagnetically aligned moments. The Eu2 (2a) Wyckoff sites lie in the center of each five member Eu ``stick", with Eu3 (4g) sites next, then Eu1 (4g) at the ends. These sticks are aligned antiferromagnetically relative to each other so there is no net moment along $\bm{\hat{b}}$. This structure agrees well with the lowest energy configuration recently reported by Crivillero $\textit{et\ al.}$ \cite{crivillero2022surface}. In a homogeneous model, the 7 K transition maintains $\bm{k}_1=(000)$ antiferromagnetism along $\bm{\hat{b}}$ while adding a $\bm{k}_2=\left(00\frac{1}{2}\right)$ component of the moment along $\bm{\hat{a}}$ on all Wyckoff sites. The principal order parameter is at the Eu1 or Eu2 site, proceeding through irrep 3.

Future work should experimentally further examine and perhaps modify our hypothesized magnetic structures (e.g., as a function of applied field as discussed in \cite{crivillero2023magnetic}), investigate in detail the interactions leading to the two magnetic phases, and develop a thorough, quantitative understanding of the relationship between magnetic and transport properties. This will contribute to knowledge of colossal magnetoresistance in antiferromagnetic $4f$-electron systems and has potential applications in spintronics technology. Our results also shed light on the magnetic properties of axion insulator candidates realized by substitution of the Eu$_5$In$_2$Sb$_6$ parent compound.

\begin{acknowledgments}
This research was conducted at the Institute for Quantum Matter, an Energy Frontier Research Center funded by the U.S. Department of Energy Office of Science, Basic Energy Sciences, under Award No. DE-SC001933. This work is based on neutron experiments performed at the NIST Center for Neutron Research. The identification of any commercial product or trade name does not imply endorsement or recommendation by the National Institute of Standards and Technology. A portion of this research used resources at the Spallation Neutron Source, a DOE Office of Science User Facility operated by the Oak Ridge National Laboratory. The MPMS was funded by the National Science Foundation, Division of Materials Research, Major Research Instrumentation Program, under Award No. 1828490. CLB and VCM were supported by the Gordon and Betty Moore foundation EPIQS program under Grant No. GBMF9456.
\end{acknowledgments}

\appendix

\section{\label{sec:hcapp}Heat Capacity}
\begin{figure}
\includegraphics[scale=1.0]{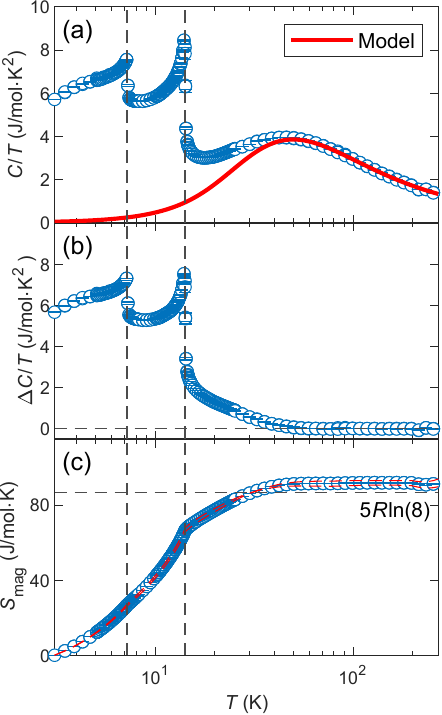}
\caption{\label{fig:hcAnharm} (a-c) Specific heat measurements on part of the $\rm Eu_5In_2Sb_6$ crystal used for neutron diffraction on BT7. (a) Debye fit with an anharmonic correction as given in Equation 1. $\Theta_{\rm{D}} = 175(5)\ \mathrm{K}$. (b) An estimate of the magnetic heat capacity determined by subtracting our calculated Debye contribution from the observed heat capacity. The magnetic contribution only declines to 0 J/mol$\cdot$K$^2$ beyond $50$ K. (c) Magnetic entropy saturates near $5 R \log{8}$ J/mol$\cdot$K (indicated by the horizontal dashed line), the expected value for a formula unit with five $S=7/2$ Eu$^{2+}$ ions.}
\end{figure}

We fit our heat capacity data using the Debye interpolation scheme while minimizing a $\chi^2_{\rm{r}}$ statistic. The data were fit from 70 K to 200 K, which is below the temperature at which Apiezon N Grease freezes but above magnetic contributions to the Eu$_5$In$_2$Sb$_6$ heat capacity (\cite{Rosa2020} and our own fitting). To estimate error bars on the fit parameters ($\gamma$ and $\Theta_{\rm{D}}$), each parameter was varied while fitting the remaining parameters until the resulting goodness-of-fit exceeded a threshold $\chi^2_{\rm{r}}$. According to the Dulong-Petit law, the heat capacity is expected to approach $3nR=324$ J/mol$\cdot$K in the high-temperature limit where $n=13$ is the number of atoms in the primitive cell. The experimental data exceeds this limit by 10\%. Experimental factors such as grease being transferred to the sample during mounting might account for this or anharmonic effects beyond the Debye model. Because the heat capacity of the grease is much larger than the heat capacity of the sample at high-temperatures, a small amount of grease transferred to the sample has a dramatic effect on the inferred sample specific heat capacity at high temperatures. In the analysis reported in the main text a scale factor $(1+\alpha)$ was applied to the measured addenda heat capacity prior to subtraction. Enforcing thus the Dulong Petit limit yields $\alpha=1.4(4)\%$ extra grease. However, the deviation from the Dulong-Petit limit can also be captured by the anharmonic model developed in \cite{stern1958} for body-centered-cubic lattices involving only nearest neighbor interactions (Fig.~\ref{fig:hcAnharm}). In that model the heat capacity is given by:
\begin{eqnarray}
\nonumber C_{(\mathrm{v,Deb})} = 9 R \left( \frac{T}{\theta_{\rm{D}}} \right)^3 \int_0^{ \frac{\theta_{\rm{D}}}{T}} \frac{x^4 e^x}{(e^{x}-1)^2} dx \\
\times \left( 1 + A T \right).
\end{eqnarray}

$C_{(\mathrm{v,Deb})}$ is the lattice contribution to the specific heat per atom in the unit cell and $A$ is the linear anharmonic correction. Despite Eu$_5$In$_2$Sb$_6$ being more complicated than is assumed in the above model, we find a linear coefficient $A = 5(1) \times 10^{-4}$ K$^{-1}$ which is near the theoretical value for Na reported in \cite{stern1958} ($A = 2.34 \times 10^{-4}$ K$^{-1}$).

The difference between the Debye heat capacity and the observed heat capacity then gives an estimate for the magnetic heat capacity.
\begin{equation}
C_{\rm{p}}(T) = n C_{(\mathrm{p,Deb})} + C_{\mathrm{mag}}(T)
\end{equation}

where $n$ is the number of atoms per formula unit. This magnetic heat capacity is integrated for the magnetic entropy
\begin{eqnarray}
S_{\mathrm{mag}}(T) = \int_0^T \frac{C_{\mathrm{mag}}(T')}{T'} dT'.
\end{eqnarray}

To integrate our discrete data we used trapezoidal numerical integration. In a primitive cell with five Eu$^{2+}$ atoms, the expected magnetic entropy is given by $S_{\mathrm{mag}} = N R \log{(2S+1)}=5 R \log{8}$. For the grease corrected fitting we find the entropy is within 2\% of the expected result for $S=7/2$ Eu$^{2+}$ ions. For the anharmonic fitting, we find the entropy is within 7\% of the expected result. By comparison, the error bar on the sample mass measurement is 2\%. NB there is additional uncertainty in the fitted parameters associated with the choice of fitting range for the Debye model.

\section{\label{sec:mag}Magnetization}
\begin{figure}
\includegraphics[scale=0.6]{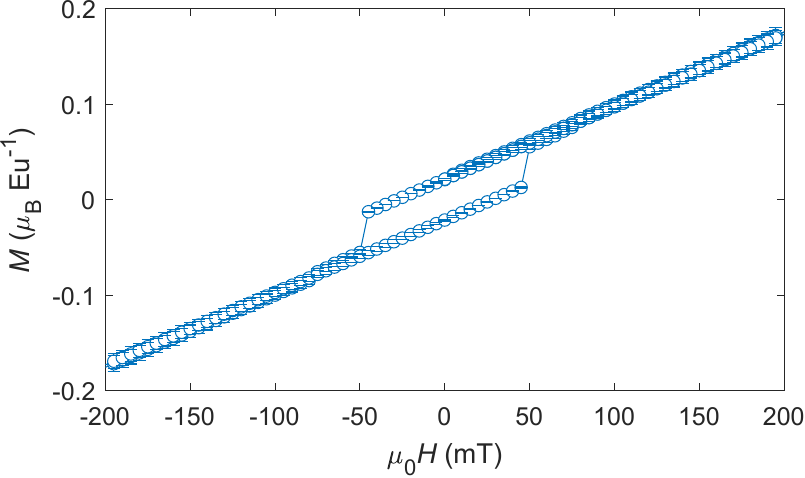}
\caption{\label{fig:MvH} Magnetization versus field applied along $\bm{\hat{a}}$ at $T = 10$~K. The sample is the 16.0(1) mg crystal from Figure~\ref{fig:overplot}f. We find a remnant magnetization corresponding to 0.021(1) $\mu\mathrm{_B/Eu}$. Our neutron diffraction is insensitive to this small ferromagnetic component of the moment along $\bm{\hat{a}}$. (Note that $1~\mu_{\mathrm{B}} = 9.274 \times 10^{-24}~\mathrm{J~T^{-1}}$.)}
\end{figure}

Magnetization along the $\bm{\hat{a}}$ axis was measured in a 7 T-MPMS on a 16.0(1) mg sample mounted with GE varnish to a quartz cylinder placed inside a brass sample holder. Magnetization along the $\bm{\hat{b}}$ axis was also measured in the MPMS on the same 17.2(1) mg sample (some mass was lost between measurements) mounted with GE varnish to a quartz cylinder placed inside a brass sample holder. Magnetization along the $\bm{\hat{c}}$ axis was measured in a 14 T-PPMS with the VSM option on a different 1.6(1) mg sample mounted with GE varnish to a quartz sample holder. All alignments were confirmed by comparing Laue X-ray backscattering spectra to the simulated patterns from QLaue \cite{qlaue}. Other than the modest increase along $\bm{\hat{b}}$ at $T_{\mathrm{N2}}$, for which we don't observe hysteresis in $M$ versus $H$ at $T=2$ K and $T=10$ K, anomalies in the susceptibility at magnetic transitions intrinsic to Eu$_5$In$_2$Sb$_6$ match those observed in \cite{Rosa2020} (Fig.~\ref{fig:overplot}f, Fig.~\ref{fig:MvH}).

\section{\label{sec:neu}Neutron Diffraction}
On TOPAZ, a 59.65 mg sample was wrapped in a thin layer of aluminum foil. Its dimensions were approximately 3$\times$3$\times$1 mm$^3$ with $\bm{\hat{b}}$ along the shortest direction. The TOPAZ data were analyzed using Mantid. For the SPINS experiment, an 80 mg crystal aligned in the $(hk0)$ scattering plane by Laue X-ray backscattering was mounted on an aluminum sample holder. For the BT7 experiment, two 30 mg crystals were aligned in the $(0kl)$ and $(h0l)$ scattering planes and co-mounted on an aluminum holder. The latter three samples were cooled using a helium flow cryostat with a 1.5 K base temperature. The neutron energy was $E_{\rm{i}}=E_{\rm{f}}=5$ meV on SPINS and 35~meV for most scans on BT7 (exceptions mentioned in text). Sets of reflections for magnetic refinements were obtained as rocking scans. The structure factors were derived from these rocking scans in MATLAB. Poor peaks located near aluminum powder lines were excluded. The most prominent nuclear, $\bm{k_2}=\left(00\frac{1}{2}\right)$, $T=10$ K $\bm{k_1}=(000)$, and $T=1.5$ K $\bm{k_1}=(000)$ peaks from each experiment were fit to Gaussian peaks with a flat background in order to obtain fixed values of the full width at half maximum (fwhm) for subsequent fitting. The four fit parameters are $I_0$, $I_{\rm{G}}$, $x_0$, and $W$ where $I_0$ is the flat background intensity, $I_{\rm{G}}$ is the Gaussian area (NB this is an integrated intensity), $x_0$ is the peak a3 position, and $W$ is the Gaussian fwhm. With the fwhm fixed by these fits to prominent peaks, all peaks were then fit. Error bars on the three fit parameters $I_0$, $I_{\rm{G}}$, and $x_0$ were determined by continuously varying one of the parameters and fitting the others until threshold $\chi^2_{\rm{r}}$ values were exceeded, then taking half the difference between the upper and lower parameter values. Rocking scan integrated intensities were determined by fits in MATLAB and converted into a full d$^3{\bm{Q}}$ integral over the Bragg peak, which is proportional to the structure factor, using ResLib with the collimation configuration and $E_{\rm{i}}$ as inputs. Thus the experimental structure factors were $F^2 = \sqrt{\frac{M_{22}}{2 \pi}} \frac{Q I_{\rm{G}}}{R_0}$ where $R_0$ and $M_{22}$ are the normalization prefactor and resolution matrix component, respectively, from ResLib and $Q$ is the scattering vector. The ratio between the largest and smallest absorption correction factor for reflections from the SPINS ($\mu_{\mathrm{abs}} = 12\ \mathrm{mm}^{-1}$) and BT7 ($\mu_{\mathrm{abs}} = 5\ \mathrm{mm}^{-1}$) experiments were 18.8 and 4.0, respectively. These short absorption lengths imply that multiple scattering can be neglected. Squared structure factors, uncertainties, and reflection indices were written to .int files for refinement of the magnetic structure in FullProf.
Refinements uncorrected for absorption had the $\bm{k}_2=\left(00\frac{1}{2}\right)$ $\bm{\hat{a}}$ components given by $ 7(1)~\mu_{\mathrm{B}}, ~3(6)~\mu_{\mathrm{B}}, ~\mathrm{and}~5(1)~\mu_{\mathrm{B}} $ for Eu1, Eu2, and Eu3, respectively. The $\bm{k}_1=(000)$ $\bm{\hat{b}}$ components were $ 2(2)~\mu_{\mathrm{B}}, ~6(3)~\mu_{\mathrm{B}}, ~\mathrm{and}~5(1)~\mu_{\mathrm{B}} $. Given the large error bars, these values do not deviate significantly from the absorption-corrected refinement where the $\bm{\hat{a}}$ components were refined within a window derived from the uncertainty of the refined $\bm{\hat{b}}$ component. Magnetic exchange energy calculations using density functional theory took the uncorrected components as inputs, with the assumption of a $7~\mu_{\mathrm{B}}$ net moment in the low-$T$ limit.

\section{\label{sec:strCalc}Magnetic Structure Factor Calculations}
Here we present calculations involving the vector structure factor, defined as:
\begin{eqnarray}
\bm{F}=\sum_d \bm{\sigma}_d \exp(i \bm{q} \cdot \bm{d}), \quad \bm{q}=h \bm{a}^*+k \bm{b}^*+l \bm{c}^*, \quad \nonumber\\ \bm{d}=x \bm{a}+y \bm{b}+z \bm{c}, \quad \bm{\sigma}_d = \sum_{i} \eta_i \bm{\sigma}_{i} \quad
\end{eqnarray}

For a given Wyckoff position, the vector structure factor $\bm{F}$ is a sum over $N$ contributions $\bm{\sigma}_{d}$ from each atomic site $d$ belonging to the $N$-fold degenerate Wyckoff position. $\bm{\sigma}_{i}$ is a set of $N$ basis vectors from the chosen irrep (Table~\ref{tab:irreps}) and $\eta_i$ is the set's mixing coefficient. In the notation of the main text, $\Gamma_3(\eta_1,\eta_2)$ refers to a magnetic configuration at some Wyckoff position whose contribution to the structure factor $\bm{F}$ is given by (sets of) basis vectors $\bm{\sigma}_{1},~\bm{\sigma}_{2}$ (found in Table~\ref{tab:irreps}) with mixing coefficients $\eta_1,~\eta_2$, respectively. Subscripts $i$ belonging to $\bm{\sigma}_{i}$ below indicate the atomic coordinates from the respective Wyckoff site (Table~\ref{tab:strFact}).
\begin{table}
\caption{\label{tab:strFact} Atomic positions, where the positions of the degenerate sites may be obtained from the given set of atomic coordinates by applying symmetry operators \cite{Park2002}}.
\begin{ruledtabular}
\begin{tabular}{|c | c || c | c | c|}
\textrm{Site}&
\textrm{Wyckoff Position}&
\textrm{x}&
\textrm{y}&
\textrm{z}\\
\colrule
Eu1 & 4g & .32749 & .01894 & 0 \\ 
\hline
Eu2 & 2a & 0 & 0 & 0 \\
\hline
Eu3 & 4g & .9128 & .7505 & 0 \\
\end{tabular}
\end{ruledtabular}
\end{table}

For a general reflection $(hkl)$ in phase 1 and Wyckoff position 2a, we find $\bm{F}_{(hkl), \mathrm{2a}} = \bm{\sigma}_1 + \bm{\sigma}_2 \exp{i \pi (h+k)}$. The contribution of the 4g sites is given by:
\begin{align*}
    \bm{F}_{(hkl), \mathrm{4g}} &= 2 (\bm{\sigma}_1 + \bm{\sigma}_2) \cos{2 \pi (hx+ky)}\\
    &\quad + 2 i (\bm{\sigma}_1 - \bm{\sigma}_2) \sin{2 \pi (hx+ky)}\\ 
    &\quad + 2 \exp{i \pi (h+k)} \left[ (\bm{\sigma}_3 + \bm{\sigma}_4) \cos{2 \pi (-hx+ky)}\right. \\
    &\left. \quad + i (\bm{\sigma}_3 - \bm{\sigma}_4) \sin{2 \pi (-hx+ky)} \right]
\end{align*}

Considering two of the reflections from our neutron diffraction temperature scans, we find $\bm{F}_{(100), \mathrm{2a}} = \bm{F}_{(011), \mathrm{2a}} = \bm{\sigma}_1 - \bm{\sigma}_2$, $\bm{F}_{(100), \mathrm{4g}} = 2 (\bm{\sigma}_1 + \bm{\sigma}_2 - \bm{\sigma}_3 - \bm{\sigma}_4) \cos{2 \pi x} + 2 i (\bm{\sigma}_1 - \bm{\sigma}_2 + \bm{\sigma}_3 - \bm{\sigma}_4) \sin{2 \pi x}$, and $\bm{F}_{(011), \mathrm{4g}} = 2 (\bm{\sigma}_1 + \bm{\sigma}_2 - \bm{\sigma}_3 - \bm{\sigma}_4) \cos{2 \pi y} + 2 i (\bm{\sigma}_1 - \bm{\sigma}_2 - \bm{\sigma}_3 + \bm{\sigma}_4) \sin{2 \pi y}$. Taking the structure to be a collinear antiferromagnet, we can calculate the structure factor for each possible configuration explicitly. Some of the vector structure factors may be further simplified by making the approximation that $x_1=0.3279 \approx \frac{1}{3}, y_1=0.01894 \approx 0$ for Eu1 and $y_3=0.7505 \approx \frac{3}{4}$ for Eu3. This approximation is good for the small wave vector transfers of our experiment and, when compared with the observed (100) and (011) reflections, leads to the $\Gamma_3$ irrep with Eu3 ordering.

Table~\ref{tab:pref} lists the factors from the differential cross section $\frac{\partial^2 \sigma}{\partial \Omega \partial E}$ which account for polarization effects by removing the component of the vector structure factor parallel to the scattering vector \cite{squires1996introduction}. This is given by $I=|\bm{F}|^2 - |\bm{\hat{Q}} \cdot \bm{F}|^2$. Mixing coefficients are given by $m_{i x}$ with $i$ giving the Wyckoff position and $x$ giving direction of the corresponding basis vector.

\begin{table*}
\caption{\label{tab:pref} Prefactors of the scattering cross section derived from the component of the vector structure factor perpendicular to the scattering vector. Numerical subscripts indicate the Wyckoff position while Latin characters indicate the crystallographic direction of the basis vector. This is an approximation in which the exact coordinates $x_1 = 0.32749$ and $y_3 = 0.7505$ were taken to $x_1 = \frac{1}{3}$ and $y_3 = \frac{3}{4}$. Coefficients were rounded to the nearest two decimal places.}
\begin{ruledtabular}
\begin{tabular}{||c||c|c|c||}
\textrm{$\Gamma$}&
\textrm{(100)}&
\textrm{(011)}&
\textrm{$\left( 0 0 \frac{1}{2} \right)$}\\
\colrule
 1 & \makecell{ $ 4 ( m_{1c}^2 - 2 m_{1c} m_{2c} + m_{2c}^2 $ \\ $ - 3.41 m_{1c} m_{3c} + 3.41 m_{2c} m_{3c} + 2.91 m_{3c}^2 ) $ } & $ 7.89 ( m_{1c}^2 + 1.01 m_{1c} m_{2c} + 0.25 m_{2c}^2 )$ & 0 \\
\hline
 2 & $12 ( m_{1b}^2 - 1.20 m_{1b} m_{3b} + 0.36 m_{3b}^2 ) $ & $ 0.23 ( m_{1a}^2 - 16.85 m_{1a} m_{3a} + 70.95 m_{3a}^2 ) $ & 0 \\
\hline
 3 & \makecell{ $ 4 ( m_{1b}^2 - 2 m_{1b} m_{2b} + m_{2b}^2$ \\ $ - 3.41 m_{1b} m_{3b} + 3.41 m_{2b} m_{3b} + 2.91 m_{3b}^2)$ } & $7.89 ( m_{1b}^2 + 1.01 m_{1b} m_{2b} + 0.25 m_{2b}^2 ) $ & $ 4 ( 2 m_{1a} + m_{2a} + 2 m_{3a} )^2 $ \\
\hline
 4 & $ 12 ( m_{1c}^2 - 1.20 m_{1c} m_{3c} + 0.36 m_{3c}^2 ) $ & 0 & 0 \\
\hline
 5 & 0 & $ 15.77 ( m_{1a}^2 + 1.01 m_{1a} m_{2a} + 0.25 m_{2a}^2 ) $ & $ 4 ( 2 m_{1b} + m_{2b} + 2 m_{3b} )^2 $ \\
\hline
 6 & 0 & $0.11 ( m_{1c}^2 - 16.85 m_{1c} m_{3c} + 70.95 m_{3c}^2 ) $ & 0 \\
 \hline
 7 & 0 & 0 & 0 \\
 \hline
 8 & 0 & $ 0.11 ( m_{1b}^2 - 16.85 m_{1b} m_{3b} + 70.95 m_{3b}^2 ) $ & 0 \\
\end{tabular}
\end{ruledtabular}
\end{table*}

\section{\label{sec:land}Secondary Order Parameters}
Regarding the phase transitions, it is understood that primary order parameters (associated with some irrep) may in general be accompanied by secondary order parameters (perhaps associated with a different irrep). But for space group $Pbam$ (with propagation vector $\bm{k}_1=(000)$ or $\bm{k}_2=\left(00\frac{1}{2}\right)$), the secondary order parameters can only be associated with the identity irrep \cite{stokes1991}. We therefore expect all magnetic order parameters sharing a critical temperature are associated with the same irrep. We consider at most three order parameters associated with a single irrep composed of axial basis vectors, corresponding to magnetic order at the three Eu Wyckoff sites.

\section{\label{sec:corr}Correlation Length}
Rocking scans were fit to a flat background and Voigt function, defined as the convolution of a Gaussian and a Lorentzian. The width $\sigma$ of the Gaussian was fixed to be the average of our brightest paramagnetic widths when fit to Gaussians. No systematic dependence of the widths on $Q$ was observed. Correlation lengths were determined by $\xi = \frac{2}{\gamma}$ where $\gamma$ is the half width at half maximum (hwhm) of the Lorentzian component of the Voigt function. Our Voigt function was integrated numerically using MATLAB's integral() function. Error bars for the hwhm were first computed by varying the hwhm and refitting the background, Voigt prefactor, Voigt center position and Gaussian width until a threshold reduced chi-squared $\chi^2_{\mathrm{r,th}} = \chi^2_{\mathrm{r,best}} \left( 1 + \frac{1}{\nu} \right) $ was exceeded. Here $\nu$ is the number of degrees of freedom, equal to the number of data points minus the number of fit parameters. Because the upper and lower bounds are asymmetric about the best fit $\chi^2_{\mathrm{r,best}}$, we individually considered each bound. We found correlation lengths in excess of 1000 \AA. Visual inspection of the fits suggested they were in fact resolution-limited, consistent with significant variation among the Gaussian widths. We therefore report lower bounds determined from a 20\% increase in $\chi_{\rm{r}}^2$, at which the deviation from the fit becomes visibly worse.

\section{\label{sec:het}Heterogeneous Model}
Here we consider several possibilities for how these phases occupy the volume of the crystal, how they compete or coexist with each other. We expect other experimental techniques will be useful in elucidating the specific nature of each phase \cite{jal2022}. There are at least four possibilities, each of which may be classified as either homogeneous or heterogeneous depending on whether or not the phases occupy the same volume. 1) Both transitions are second-order phase transitions. The 7 K transition simply adds the $\bm{k}_2=\left(00\frac{1}{2}\right)$ propagation vector to phase 1 forming a homogeneous crystal. 2) Different volume fractions have distinct second-order phase transitions (either at 14 K or at 7 K), perhaps due to chemical differences. Each transition is from a paramagnetic state to an ordered state. This heterogeneous model is inconsistent with the reduction in (011) intensity as we would not expect significant competition between the two isolated phases. 3) Below 7 K there's an an incomplete second-order transition to a single-$\bm{k}$ structure. Here, part of the $\bm{k}_1=(000)$ phase transitions to a $\bm{k}_2=\left(00\frac{1}{2}\right)$ phase below 7 K. The 14 K transition is from the paramagnetic state while the 7 K transition is from the ordered state. This is a heterogeneous model which again could result from chemical differences. 4) Incomplete conversion via a first-order phase transition to a heterogeneous structure with $\bm{k}_1=(000)$ and $\bm{k}_2=\left(00\frac{1}{2}\right)$ phases below 7 K. But we find the 7 K transition appears to be second-order in heat capacity so we're left with models 1 and 3. We focus on model 1 but also present results for model 3.

In model 3, the crystal undergoes an incomplete second-order phase transition in which particular volumes transition from a pure $\bm{k}_1=(000)$ phase to a pure $\bm{k}_2=\left(0 0 \frac{1}{2}\right)$ phase to give a heterogeneous sample. Here, the $\bm{k}_2=\left(0 0 \frac{1}{2}\right)$ volume is in general allowed to possess both $\bm{\hat{a}}$ and $\bm{\hat{b}}$ components without giving an oscillating moment size upon approaching low-$T$. There is, however, little anomaly seen in the susceptibility along $\bm{\hat{b}}$ at the 7 K transition so it appears the staggered moment of the $\bm{k}_2=\left(0 0 \frac{1}{2}\right)$ phase is still primarily along $\bm{\hat{a}}$. We refine structures for both phases, each of which has the saturated moment (Fig.~\ref{fig:refinement}c,d). Unlike the homogeneous case, all Wyckoff positions in a given phase must have moments with identical magnitudes and directions. This is somewhat compensated by the additional degree of freedom that allows for different relative volume fractions of the two phases. We find the refinements for the heterogeneous model are quite similar to the homogeneous model (Fig.~\ref{fig:refinement}h). Based on the refined scale factors at 1.5 K, we find the phase 1 volume fraction to be $70(38)\%$ and the phase 2 volume fraction to be $40(14)\%$ under the constraint that the sum of volume fractions cannot exceed 100\%. The magnetic space group of this second pure phase is 62.451 in the Belov-Neronova-Smirnova (BNS) notation, in which the unit cell is generally not the paramagnetic unit cell \cite{stokes}. This is 55.9.449 according to the Opechowski-Guccione (OG) notation, in which the unit cell is the paramagnetic unit cell. The structure corresponds to the A-type antiferromagnetic configuration proposed in \cite{Rosa2020}.

On the one hand, energy scales from DFT appear to favor the two distinct single-$\bm{k}$ phases composing the heterogeneous model. Figure~\ref{fig:energyscales}b shows the experimentally determined low-temperature homogeneous state lies near a maximum in energy between two minima. The minima correspond to the two heterogeneous phases. On the other hand, in a heterogeneous sample we would expect different syntheses to produce different chemical phase fractions. The sharp peaks in heat capacity and consistency between heat capacity measurements are not expected for this model. Furthermore, the (011) temperature scan is not consistent with the heterogeneous picture. Specifically the (011) intensity was observed to sharply decline at 7 K. Influence of the $\bm{k}_2=\left(00\frac{1}{2}\right)$ transition on the $\bm{k}_1=(000)$ structure suggests the k-vectors are not separated in space as is assumed in the heterogeneous model. Rather the low-$T$ state consists of a single homogeneous magnetic phase in which a shift of intensity from $\bm{k}_1=(000)$ to $\bm{k}_2=\left(00\frac{1}{2}\right)$ reflections is simply a consequence of the reorientation of magnetic moments on each atomic site.

\nocite{*}

\bibliography{Eu5In2Sb6}

\end{document}